\documentstyle[sprocl]{article}
\newcommand{\bd}{\begin{document}}
\newcommand{\ed}{\end{document}}
\newcommand{\bc}{\begin{center}}
\newcommand{\ec}{\end{center}}
\newcommand{\be}{\begin{eqnarray}}
\newcommand{\ee}{\end{eqnarray}}
\newcommand{\eqn}{\global\def\theequation}
\newcommand{\sw}{sin^2 \theta_W}
\renewcommand{\thefootnote}{\fnsymbol{footnote}}

\begin{document}
\title{
How to Observe the Interference Effects of Top Quark Polarizations at 
Tevatron \footnote[1] {Contribute to Lake Louise Winter Institute
(1996).}?} 
\author{Darwin Chang}

\address{Department of Physics, National Tsing Hua University\\
Hsinchu, 30043, Taiwan, Republic of China}


\maketitle\abstracts{ Using a simple analytic expression for 
$q \bar{q}, g g \rightarrow t \bar{t} \rightarrow b W^+ \bar{b} W^- 
\rightarrow b \bar{l} \nu_l \bar{b} l' \bar{\nu_{l'}}$ 
with the interference effects due to the polarizations of the $t$ and 
$\bar{t}$, we demonstrate how the effects can be measured at Tevatron at 
$3\sigma$ level. 
}


With the upgrade of Tevatron and the improvements of its two collider  
detectors, the data
related to the production of the top quark are going to accumulate very
quickly.  Since top quark decays very rapidly after its production, it is 
expected that the spin information of the top quarks is preserved in the 
decay process.  For lighter quark, this spin information is always smeared 
by the hadronization effect.  Therefore, if such effect can be observed it 
will be the first time that one can observe the spin of a bare quark 
directly.  However, it is 
challenging to investigate how one can determine the various
detail properties of top quark in the complex hadronic collider environment.

Here we report our investigate an interesting physical consequence of the 
fact that the top quarks that are produced and decayed are supposed to be 
spin $1/2$ particles\cite{clt,cls}.  
One of the important effect of the polarizations of 
unstable particles are that the different polarized intermediate states 
can interfere.  
In this sense, the observation of interference effects provides a 
unique possibility of direct observation of the spin of a quark. 

We use a simple analytic result\cite{clt} for the differential cross 
section of $q \bar{q}$ and $g g \rightarrow 
t \bar{t} \rightarrow b W^+ \bar{b} W^- \rightarrow b \bar{l} \bar{\nu_l}
\bar{b} l' \nu_{l'}$ based on an analytic helicity technique
developed in ref. \cite{leesc}.
The decays of $W$ bosons and top quarks are taken into account in the narrow 
width approximation. 
The contribution due to off-shell 
top quarks or off-shell $W$ bosons are negligible.
The interference effects discussed here was also considered before in
ref. \cite{kane}, however, it was studied only numerically and was 
done in a rougher approximation.  
The analytic expressions obtained here can also be easily adapted to the 
leptonic colliders\cite{ccls}.  We shall also demonstrate that such 
interference effect can be detected at $3 \sigma$ level at Tevatron with 
Main Injector if one uses proper observables\cite{cls}.

The polarization density matrix for the process can be split into two 
main sections and written as
\begin{eqnarray}
{\it P} &=& {1\over N_{s}} \sum_{
\lambda_1, \lambda_2, \lambda_3, \lambda_3', \lambda_4, \lambda_4'}
{\it P}(i_{\lambda_1\lambda_1}\bar{i}_{\lambda_2\lambda_2}\rightarrow 
t_{\lambda_3\lambda_3'}\bar{t}_{\lambda_4\lambda_4'})
\left | \Pi_{t}(r_1) \right |^2
\left | \Pi_{\bar{t}}(r_2) \right |^2
\nonumber\\&&\qquad\qquad\qquad\qquad\times
{\it P}(t_{\lambda_3\lambda_3'}\bar{t}_{\lambda_4\lambda_4'}\rightarrow 
b\bar{b}l^+l^-\nu\bar{\nu}) .
\end{eqnarray}
Here the initial the final polarizations are summed over.
$N_{s} = 4 (N^2 - 1)^2$ for $gg$ initial states; and
$N_{s} = 4 N^2$ for $q\bar{q}$ initial states.
$\Pi_{q}(r_i)=-i(r_i^2-m_{q}^2)^{-1}$ 
represents the polarization independent components of the top quark 
propagators.  The helicity informations are included in the remaining 
density matrices. 
When no subscript is included for a given particle with spin in $P$, 
it implies that its helicity is summed over.
The polarization density matrix
$P(i_{\lambda_1\lambda_1}\bar{i}_{\lambda_2\lambda_2}\rightarrow 
t_{\lambda_3\lambda_3'}\bar{t}_{\lambda_4\lambda_4'})$
represents the production of the top quark pair via the processes
$qq\rightarrow t\bar{t}$ or $gg\rightarrow t\bar{t}$. 
$P(t_{\lambda_3\lambda_3'}\bar{t}_{\lambda_4\lambda_4'}
\rightarrow b\bar{b}ll\nu\nu)$ 
is the polarization density matrix 
for the decay of the top quark pair into $b$ quarks and leptons.  
This density matrix can be further splitted into the product of 
the decay density matrix of
$t \bar{t}$ into $W^+ W^-$ bosons and $b \bar{b}$ pairs, and the decay
density matrix of $W^+ W^-$ boson pair into a pair of fermions each. 

The polarized density matrix of the process
$t_{\lambda_3\lambda_3'}\bar{t}_{\lambda_4\lambda_4'}
\rightarrow b\bar{b}ll\nu\nu$ can be written as 
\begin{eqnarray}
P&(&t_{\lambda_3\lambda_3'}\bar{t}_{\lambda_4\lambda_4'}\rightarrow 
b\bar{b}l^+l^-\nu\bar{\nu}) =
\sum_{\lambda_5, \lambda_5', \lambda_6, \lambda_6'}
P(t_{\lambda_3\lambda_3'}\rightarrow W^+_{\lambda_5\lambda_5'} b) 
P(\bar{t}_{\lambda_4\lambda_4'}\rightarrow 
W^-_{\lambda_6\lambda_6'} \bar{b})
\nonumber\\&&\times
\left | \Pi_{W}(p_1) \right |^2
\left | \Pi_{W}(p_2) \right |^2
P(W^+_{\lambda_5\lambda_5'} \rightarrow l^+ \bar{\nu})
P(W^-_{\lambda_6\lambda_6'} \rightarrow l^- \nu).
\end{eqnarray}
Simple analytic expressions for all the polarization density matrices 
appeared in the first hand side of above equation can be found in 
Ref.\cite{clt}.
These expressions can be combined together with 
the narrow width approximation for the $W$ boson propagators and obtain
\begin{eqnarray}
P&(&t_{\lambda_3\lambda_3'}\bar{t}_{\lambda_4\lambda_4'}\rightarrow
b\bar{b}ll\nu\nu)
= 
16 m_t^2 \left (
{ e^4 \left | V_{tb}\right |^2 \over 2 \sin^4\theta_w } \right )^2
{\pi^2\delta \left ( p^2_{1} - m^2_W \right )
\delta \left ( p^2_{2} - m^2_W \right )
\over ( M_W \Gamma_W )^2 }
\nonumber\\&&\times 
(l_{1} \cdot e_{1}^a)(k_{1}\cdot q_{1})
(l_{2} \cdot e_{2}^b)(k_{2}\cdot q_{2})
\sigma_{a(\lambda_3\lambda_3')}
 \otimes \bar{\sigma}_{b(\lambda_4\lambda_4')}.
\end{eqnarray}
\begin{eqnarray}
P&(&t_{\lambda_3\lambda_3'}\bar{t}_{\lambda_4\lambda_4'}\rightarrow 
b\bar{b}l^+l^-\nu\bar{\nu}) 
=
16 m_t^2 \left (
{ e^4 \left | V_{tb}\right |^2 \over 2 \sin^4\theta_w } \right )^2
{\pi^2\delta \left ( p^2_{1} - m^2_W \right )
\delta \left ( p^2_{2} - m^2_W \right )
\over ( M_W \Gamma_W )^2 }
\nonumber\\&&\times
(l_{1} \cdot e_{1}^a)(k_{1}\cdot q_{1})
(l_{2} \cdot e_{2}^b)(k_{2}\cdot q_{2}) 
\sigma_{a(\lambda_3\lambda_3')}
 \otimes \bar{\sigma}_{b(\lambda_4\lambda_4')}.
\end{eqnarray}

The helicities of final state fermions are summed over. 
We shall use four vectors $k_i$, $l_i$ and $q_i$ to represent the 
momenta of the neutrinos, the charged leptons and the b quarks respectively.
The subscript 1 is used to label the momenta related to the decay 
products of the top quark, while the subscript 2 is used for those  
related to the decay of the anti-top quark. 
$\epsilon_{1\mu}(\lambda_5)$ 
represents the polarization vectors of $W^+$ with helicity 
$\lambda_5$, while 
$\epsilon_{2\nu}(\lambda_6)$
represents those of the $W^-$ with helicity 
$\lambda_6$.
In our notation, the information about the top helicity is carried by
the Pauli matrix $\sigma_i$ and $\sigma_0 (= 1)$ 
with component $(\lambda_3\lambda_3')$.
 For the anti-top, we used the conjugate matrix 
$\bar{\sigma}_a = \sigma_2 \sigma_a \sigma_2 $
with component $(\lambda_4\lambda_4')$.
The tensor $e^a (r_i)$, $(a = 0,1,2,3)$, were introduced in 
ref.\cite{leesc} to evaluate the product of fermion wave functions, 
$u(r,\lambda)\bar{u}(r,\lambda')$, 
in the amplitude squared for the cross section.  

With this, one can then proceed to glue together the density matrix for 
the production, which has been given in Ref.\cite{leesc}, and those
for the decays of the top quarks. 
 For the top and anti-top propagators, we shall use also 
the narrow width approximation. 
These provide another two delta functions.
After that, we can integrate out all the delta functions in the phase 
space for 6 final state particles and obtain the  
differential cross section 
\begin{eqnarray}
d\sigma_{ii} &=& {P \over 4 i_1\cdot i_2}
d_6 ( T \rightarrow k_{1} l_{1} q_{1} k_{2} l_{2} q_{2} )
\nonumber\\
&=&
{4 \pi^2 \alpha_s^2\over N} \left (
{ e^4 \left | V_{tb}\right |^2 \over 2 \sin^4\theta_w } \right )^2
(k_{1} \cdot q_{1})(k_{2}\cdot q_{2}) l_{1\alpha} l_{2\beta} 
I_{ii}^{\alpha\beta} \,
{d_2 (T\rightarrow r_{1} r_{2})\over i_{1}\cdot i_{2}}
\nonumber\\
&&\times
{d_2 (r_{1} \rightarrow p_{1} q_{1})
d_2 (r_{2} \rightarrow p_{2}  q_{2})
d_2 (p_{1} \rightarrow l_{1} k_{1})
d_2 (p_{2} \rightarrow l_{2} k_{2})\over 
\left ( M_W \Gamma_W M_t \Gamma_t \right )^2 },
\end{eqnarray}
where
$T^\mu = i^\mu_1+i^\mu_2$,
with $i_1$ and $i_2$ defined as the four-momenta of the initial partons
and $d_2 ( a \rightarrow b c)$
is the phase space for a particle with four-momentum $a$ 
decaying into particles with four-momenta $b$ and $c$.
$N$ is the number of color and the indices
$ii$ designate the initial particles.
 For the $g g \rightarrow t \bar{t}$ case, we have the result 
\begin{eqnarray}
I_{gg}^{\alpha\beta}
&=& \left ( x - 1 - {1\over N^2-1} \right ) \left \{  
2  r_{1}^\alpha r_{2}^\beta  \left [ {-1\over x} 
+ {1\over \gamma^2} \left ( 1- {x\over 2\gamma^2}\right )
+ 1 - {x\over 2\gamma^2} \right ] \right. 
\nonumber\\   
&&
- {1\over 2\gamma^2}\left( 1 - {x\over \gamma^2} \right )
\left (  r_{1}^\alpha T^\beta  +
 T^\alpha r_{2}^\beta  \right )
+ m^2_t \left ( {-1\over x} + {1\over \gamma^2}
\left ( 1 - {x\over 2\gamma^2} \right )
 \right )   g^{\alpha\beta} 
\nonumber\\
&&
+ \left.{1\over 4\gamma^2}
\left ( 1 - {x\over 2 \gamma^2} \right ) 
\left [ T^\alpha T^\beta 
- R^\alpha R^\beta 
-\left ( {u-t\over s} \right ) \left (
 R^\alpha T^\beta - T^\alpha R^\beta  \right ) \right ] \right \}
\end{eqnarray}
and for the $q \bar{q}\rightarrow t \bar{t}$ case
\begin{eqnarray}
I_{qq}^{\alpha\beta}
&=& \left ( {N^2-1 \over 2 N} \right ) \left \{
2 r_{1}^\alpha r_{2}^\beta  \left [ 1 
-{1\over x} \left (1 - {x\over 2\gamma^2} \right )\right ] 
-{1\over 2\gamma^2}\left ( r_{1}^\alpha T^\beta  +
 T^\alpha r_{2}^\beta \right )\right.
\nonumber\\
&&
- \left. {m^2_t\over x} \left ( 1 - {x\over 2 \gamma^2} \right )
 g^{\alpha\beta}
+ {1\over 4\gamma^2}
\left [ 
 T^\alpha T^\beta 
- R^\alpha R^\beta  
-\left ( {u-t\over s} \right ) \left (  R^\alpha T^\beta 
- T^\alpha R^\beta \right ) \right ]
\right \}.
\end{eqnarray}
The $s$, $t$ and $u$ are the Mandelstam variable,
$R^\mu = i^\mu_1-i^\mu_2$,
$\gamma^2 = s/4m_t^2$ and
$x=s^2/2(t-m_t^2)(u-m_t^2)$.

To show the spin effect of the quark, 
we shall compare with the similar calculation in which 
top quark helicities are averaged over both in the 
production density matrix and in the decay density matrix independently 
before we join them together.  We shall refer to this case as the 
"spinless case".  In the usual Monte Carlo simulation for the 
experiments, this is indeed the approximation taken by the program.
The cross section for the spinless case can be obtained by substituting
$I_{gg}$ and $I_{qq}$ with 
\begin{eqnarray}
I_{\overline{gg}}^{\alpha\beta}
&=& \left ( x - 1 - {1\over N^2-1} \right ) \left \{  
  r_{1}^\alpha r_{2}^\beta  \left [ 1 - {1\over x} 
+ {1\over \gamma^2} \left ( 1- {x\over 2\gamma^2}\right ) 
\right ] \right \}
\end{eqnarray}
and
\begin{eqnarray}
I_{\overline{qq}}^{\alpha\beta}
 &=& \left ( {N^2-1 \over 2 N} \right ) \left \{
 r_{1}^\alpha r_{2}^\beta  \left ( 1 
-{1\over x} \left (1 - {x\over 2\gamma^2} \right )\right ) 
\right \}.
\end{eqnarray}
 For the last two equation, we should emphasize that
the cross section without the spin correlation for the
top quark is the one that has been used in the literature and by most of 
the experimental event simulators. 

To measure interference effect, we need to find observables
that correlate the kinematic variables associated with the particles from 
the top quark decay and those from the anti-top quark decay.  
Intuitively, it is not obvious what is the best correlated observable 
which can probe this interference effect.  For the rest of this paper, we 
shall present our attempts in this direction.

To probe the interference effect, we shall look for observables
related to the final decay products, instead of using directly the top 
momenta which need to be reconstructed.
The easiest observables to try are those related to the two charged 
leptons (electron or muon) from the leptonic decays of two $W$'s. The 
advantage is that these particles can be easily observed in a 
hadron collider.  On the other hand, 
the dilepton decays has
lower branching ratios, which reduces the statistics.
 For this case we had investigated \cite{clt} the effect of interference 
terms on the  distribution of 
(1) their total energy $E_{l_{1}}+E_{l_{2}}$, (2) their total z-momentum 
$l_{1z}+l_{2z}$, (3) an orthogonal combination $l_{1z}-l_{2z}$, 
(4) the cosine of the angle between these two charged leptons and (5) the 
asymmetries $A_P$ to be defined later. 

Here we shall simply report the observables that we think are most promising.
We shall demonstrate that it is 
possible to observe such asymmetries at $3\sigma$ level with Tevatron II.
The lepton correlated asymmetry, $A_P$, with 
respect to a plane $P$ passing through the interaction point is defined 
as follows.  Let $\vec{l}_1$
be the momentum of the charged lepton associated with top decay  
evaluated in the top rest frame.
Similarly, let $\vec{l}_2$ be that associated with anti-top decay  
evaluated also in its own rest frame.
Then, define $A_P = (N_+ - N_-)/(N_+ + N_-)$ where 
$N_+$ is the number of 
events in which both 
$\vec{l}_1$ and $\vec{l}_2$
lies on the same side of $P$ while 
$N_-$ is the number of 
events with both 
$\vec{l}_1$ and $\vec{l}_2$
lying on the opposite side of $P$.
By choosing different $P$, one can construct different asymmetries.
These asymmetries vanish when the effects of both top spin and W spin are 
ignored (or averaged over) and 
they remain small even when the W spin effects are included.  This 
property makes them the ideal candidates for the observation of the 
$t\bar{t}$ spin correlation effects.

In order to measure the lepton correlated asymmetries it is essential to 
make full reconstruction of the top dilepton events.  We assume that 
the top mass will be well measured in Tevatron Run I and Run II.  For 
dilepton candidate events, we further assume that the missing transverse 
momentum measured is equal to the sum of the missing transverse momenta of 
the two neutrinos associated with the dilepton.  Contribution to the 
missing transverse momentum from other neutrinos in the event reduces the 
efficiency of the reconstruction and lowers the signal-to-noise ratio but 
does not spoil the observability of the asymmetries we consider here.  

The analysis of the productions and the decays of $t\bar{t}$ in 
hadronic collider with and without spin correlation effect can be found in 
the literature\cite{clt,barger} with various degrees of sophiscations.  
We shall use the simple analytic differential cross sections for  
$q \bar{q}$ and $g g \rightarrow 
t \bar{t} \rightarrow b W^+ \bar{b} W^- \rightarrow b \bar{l} \bar{\nu_l}
\bar{b} l' \nu_{l'}$ 
provided in Ref.\cite{clt}.
Using these formulas, we simulated the $t\bar{t}$ production and decay 
employing the event generator PYTHIA 5.7.  
The algorithm of our reconstruction of the top dilepton events has been 
described in detail in Ref. \cite{cls}.

In principle, one can also look for the $W$ correlated asymmetries for 
the two $W$'s from $t\bar{t}$ decays in the reconstructed top dilepton, 
lepton-jet and jet-jet events.  The predicted $W$ asymmetries can be 
easily obtained from Ref.\cite{clt}.  They are no more than a few per 
cent and about an order of magnitude smaller than some of the lepton 
correlated asymmetries.  The enhancement of the lepton correlated 
asymmetries over the $W$ correlated asymmetries\cite{cls} is one feature 
that favors the observation of top spin correlation effects through the 
dilepton channel.

In hadron colliders, $t\bar{t}$ are produced either by quark-antiquark 
($q\bar{q}$) annihilation or by gluon-gluon ($gg$) fusion.  The lepton 
correlated asymmetries for these two cases typically have opposite sign 
and with different magnitudes\cite{cls}.  
At Tevatron energy, 
the number of $q\bar{q}$ produced top events are roughly eight times that 
of the $gg$ produced events.  
This ratio decreases with energy and, at LHC energy, 
the number of $gg$ produced top events are roughly four times that 
of the $q\bar{q}$ produced events.  
For the asymmetries we considered, we found that the substantial 
asymmetries that can be observed at Tevatron become much smaller at the 
expected LHC collider energy.   
This is the result of accidental cancellation between the 
contribution of the $q\bar{q}$ production channel and that of the $gg$ 
production channel.  We have checked that the contribution due to the 
$q\bar{q}$ channel alone is indeed not so small.  
This cancellation could be a generic feature which may 
make LHC unfavorable machine to look for top spin correlation effect.  
 For the same reason, the next linear colliders should be an ideal machine 
for observing such correlation due to the absence of such cancellation.

As examples, we shall discuss the asymmetries with respect to the 
following planes defined in the $t\bar{t}$ center of mass frame: (1). 
$t\bar{t}$ production plane (defines asymmetry $A_1$); (2). the plane 
perpendicular to the production plane and contains the top (asymmetry 
$A_2$); (3). the plane perpendicular to the two previous planes (asymmetry 
$A_3$); (4). the plane normal to the beam direction (asymmetry $A_4$).  
These planes are chosen as samples to demonstrate the possibilities.  
The optimal choice will depend on the details of the detectors and 
clearly need further study\cite{cls}.  

The typical trigger for top dilepton events, namely, 
$p_T \geq 20 GeV$ for leptons, 
$p_T \geq 10 GeV$ for $b$-jet, missing transverse energy 
$\not{\!\!E}_{\perp} \geq 25 GeV$ and rapidity $|\eta| \leq 2$, was applied.
Afterwards, the reconstruction algorithm we described earlier was carried 
out.
A typical hadron calorimeter energy resolution of $70\%/\sqrt{E}$ was used 
as b - jet energy smearing and the missing transverse energy was Gaussian 
smeared with a $15\%$ standard deviation.
The effect of including contribution of other neutrinos 
in an event to the missing transverse energy 
was investigated\cite{cls}.  To isolate the effect of the bias originated 
from  
reconstruction algorithm to the asymmetries, we also studied the case 
with the event reconstruction turned on but with the trigger turned off.  
Similarly we also studied the case  
with the event reconstruction turned off but with the trigger turned on.  
The effect of energy smearing is also investigated. 
In each case, we computed the lepton correlated asymmetries, 
the neutrino correlated asymmetries and the $W$ 
correlated asymmetries with respect to the planes described earlier.  
In Table \ref{t1}, we present the measured correlated asymmetries 
$A_1, A_2, A_3$ and $A_4$
for the four cases: 
(I). trigger off, reconstruction off;
(II). trigger on, reconstruction off;
(III). trigger off, reconstruction on;
(IV). trigger on, reconstruction on;
(V). trigger on, reconstruction on and with energy smearing included. 
The corresponding asymmetries
when the top spins were ``uncorrelated'' 
(that is, their spins are summed over in their productions and 
averaged over in their decays,)
are given in brackets.
The average 
values and the standard deviations of the asymmetries were extracted 
directly from the simulated data.  When both top and $W$ spins were 
``uncorrelated'', 
as in most standard event generator packages, we verified that all the 
asymmetries vanish if the trigger and event reconstruction were turned off.
The top quark mass is taken to be $m_t =$ 176 GeV in all numerical analyses.
As one can see in the table, the asymmetries measured by charged 
leptons are generally larger than the asymmetries measured by neutrinos 
and $W$ bosons.  
The systematic effects of trigger and reconstruction are quite obvious in 
the case of neutrino asymmetries as one would expect from the large 
missing $E_{\perp}$ cut as well as the contributions from other neutrinos 
in the event.  Due to space, a detailed discussions of the 
various effects and their origins will be given elsewhere.

 From the results of these simulation, we conclude that the lepton 
correlated asymmetries arising from the $t\bar{t}$ spin correlation 
can be observed at 3$\sigma$ level with a few hundred top dilepton 
events.  
This is certainly reachable with the projected luminosity of the 
Tevatron for Run II, with improved detector resolution and acceptance 
expected for both CDF and D0 detectors upgrades and perhaps with improved 
algorithms for identifying the top dilepton events.  

There are also plenty of rooms for improvement on our analysis of the 
correlated asymmetries.  Even though trigger and reconstruction show 
little systematic effects for the lepton asymmetry, they do shift the 
neutrino and the $W$ asymmetries by non-negligible ammounts. A 
quantitative understanding of these effects may allow us to tune the 
trigger selection and reconstruction algorithm for the observation of 
the asymmetries.  It is well known that discrete ambiguities exist in 
general in the reconstruction of top dilepton events.  A detailed 
quantitative study of its effect on the asymmetries could precipitate an 
improvement on reconstruction algorithm for better efficiency and better 
signal-to-noise ratio.  
We have been quite casual in choosing the planes to define the 
asymmteries and in choosing the combinations of these asymmetries to 
measure. 
One may wonder if 
there is an optimal choice of plane or combinations of planes that can 
maximize the observability of the correlation effect.  One may also 
wonder if there are choices that will enhance or suppress the 
contribution of $q\bar{q}$ annihilation channel relative to the $gg$ 
fusion channel.

Before we conclude, we shall mention two other works\cite{mp,sw} which 
appear in the literature recently.  These two papers ignore the 
off-diagonal terms of the top production density matrix and discuss only 
the spin correlated asymmetry.  In contrary, our work addresses the more 
general density correlated asymmetry in top production.  
Only the contributions of the diagonal terms can be described as spin 
correlated.  When off-diagonal terms are included in the analysis, most 
of the time the top quarks are produced in mix states of helicity(or 
spin) variable.  As a result, our analysis includes effects that cannot be 
described as spin correlated asymmetry.  
Of course, for some observables the contributions of the off-diagonal 
density matrix may happen to be negligible.  However it is generally not 
the case. In fact the optimal case we discovered, the $A_4(l)$ asymmetry 
in the Table, cannot be described correctly by ignoring the off-diagonal 
terms.  While the spin correlated asymmetry may enjoy simpler intuitive 
understanding, it, however, does not reflect the full range of possible 
interference effects emersed in the density matrix correlation in the top 
pair prodction.

In conclusion, top spin correlation is certainly one of the most 
interesting top physics 
to be uncovered.  It can provide a direct observation of the spin $1/2$ 
character of top quark (which we have not been able to do for the lighter 
quarks) and can potentially test the $V-A$ character of the weak charged 
current associated with top.  We have clearly demonstrated the possibility 
of observing this effect at Tevatron Run II.  The fact that this effect 
may be even harder for LHC to measure should make the task more important 
for Tevatron.  A detailed account of our study will be presented 
elsewhere\cite{ccls}.

~~~~~We would like to thank Shih-Chang Lee, Paul Turcotte, Alexei Sumarokov, 
Vernon Barger, Jim Ohnemus, 
Stephen Parke and Chris Quigg for useful discussions. This work is supported
by a grant from National Science Council of Republic of China.


%
%
%

\onecolumn

\begin{table}
\caption{
The measured correlated asymmetries $A_i(l)$'s, $A_i(\nu)$'s 
and $A_i(W)$'s 
for the four cases: 
(I). trigger off, reconstruction off;
(II). trigger on, reconstruction off;
(III). trigger off, reconstruction on;
(IV). trigger on, reconstruction on;
(V). trigger on, reconstruction on with energy smearing.
The corresponding asymmetries when the top spins were ``uncorrelated'' 
are given in brackets.  All the values are in unit of percentage.
The total number of simulated events is 100000.
}
\label{t1}
\begin{tabular}{cccccc}
$$       &         $I$          &    $II$         & $III$        & $IV$       &
$V$
\\
\hline
$A_1(l)$ & $  3.99(-.11)$&$  5.90( .58)$&$  4.16(  .68)$&$  6.42( 1.94)$&$ 
5.50(1.14)$
\\
$A_2(l)$ & $-11.78( .41)$&$-11.18(-.55)$&$-10.30( -.30)$&$ -7.98(  .12)$&$
-7.58(-.06)$
\\
$A_3(l)$ & $ -9.65(-.09)$&$-8.68(-1.34)$&$ -8.00(-1.22)$&$ -6.80( -1.4)$&$
-6.64(-1.26)$
\\
$A_4(l)$ & $-20.34(
.18)$&$-17.64(-.13)$&$-16.68(-1.86)$&$-13.72(-1.66)$&$-13.30(-2.00)$
\\
$A_1(\nu)$&$  0.54(-.68)$&$  9.50(8.86)$&$  5.44( 4.70)$&$ 12.24(11.80)$&$ 
9.02( 8.48)$
\\
$A_2(\nu)$&$ -1.95(-.06)$&$   .60(1.68)$&$  1.80( 2.20)$&$  2.16( 2.42)$&$ 
2.56( 2.52)$
\\
$A_3(\nu)$&$ -1.45(-.09)$&$  3.62(4.30)$&$ -1.94(-1.54)$&$  2.64( 2.82)$&$  
.74( 1.22)$
\\
$A_4(\nu)$&$ -3.41(-.37)$&$ -1.84(-.13)$&$   .33(  .27)$&$  -.02(  .24)$&$  
.84( 1.02)$
\\
$A_1(W)$ & $   .79( .16)$&$  3.24(2.32)$&$  1.92( 1.34)$&$  3.94( 3.18)$&$ 
3.70( 2.92)$
\\
$A_2(W)$ & $ -1.98(-.51)$&$  -.48( .33)$&$ -3.34( -.12)$&$ -3.14(  .58)$&$
-2.96(  .58)$
\\
$A_3(W)$ & $ -2.70(-1.46)$&$  .56( .42)$&$ -3.70(-1.44)$&$ -3.64(-1.32)$&$
-4.60(-2.34)$
\\
$A_4(W)$ & $ -3.66( -.91)$&$-3.32(-.04)$&$ -7.12(-1.68)$&$ -6.58( -.72)$&$
-7.57(-1.62)$
\\
\end{tabular}
\end{table}

\begin{thebibliography}{99}
%
\bibitem{clt} 
D. Chang, S.-C. Lee and P. Turcotte, 
hep-ph-9508357, to appear in Chinese Journal of Physics (1996).
%
\bibitem{cls} 
D. Chang, S.-C. Lee and A. Soumarokov, Preprint 
hep-ph-9512417, to appear in Physical Review Letters.
%
\bibitem{leesc} S.C. Lee, Phys. Lett. B {\bf 189}, 461, 1987.
%
\bibitem{kane} G.L. Kane, G.A. Ladinsky, C.-P. Yuan, Phys. Rev. D {\bf 45},
124, 1992.
%
\bibitem{ccls} 
D. Chang, C.-H. Chen, S.-C. Lee and A. Soumarokov, in preparation.
%
\bibitem{barger}
R. Kleiss and W.J. Stirling,
Z. Phys. C {\bf 40} 419 (1988);
V. Barger, J. Ohnemus and R.J.N. Phillips,
Int. J. Mod. Phys. A {\bf 4} 617 (1989).
%
\bibitem{mp} 
G. Mahlon and S. Parkes, 
hep-ph-9512264, Dec. (1995).
%
\bibitem{sw} 
T. Stelzer asnd S. Willenbrock
hep-ph-9512292, Dec. (1995).
%
\end{thebibliography}
\end{document}